\documentclass[review, english, sort&compress]{elsarticle}
\usepackage{amsmath}
\usepackage{amssymb}
\usepackage{amscd}
\usepackage{graphicx}
\usepackage{adjustbox}
\usepackage{pifont}
\usepackage{natbib} 
\usepackage{geometry}
\usepackage{fleqn}
\usepackage{hyperref}
\usepackage{amsfonts}
\usepackage{makeidx}
\usepackage{lmodern}
\usepackage[T1]{fontenc}
\usepackage{float}
\usepackage{multirow}
\usepackage{textcomp}
\usepackage[english]{babel}
\usepackage[flushleft]{threeparttable}
\usepackage{epstopdf}
\usepackage{bm} 
\usepackage{lineno}    
\usepackage{xcolor}
\usepackage{pgf}
\usepackage{chngcntr}

\makeatletter
\def\ps@pprintTitle{%
 \let\@oddhead\@empty
 \let\@evenhead\@empty
 \def\@oddfoot{}%
 \let\@evenfoot\@oddfoot}
\makeatother

\makeatletter
\renewcommand\appendix{\par
  \setcounter{section}{0}%
  \setcounter{subsection}{0}%
  \setcounter{equation}{0}%
  \setcounter{table}{0}
  \setcounter{figure}{0}
  \gdef\theequation{\@Alph\c@section.\arabic{equation}}%
  \gdef\thefigure{\@Alph\c@section.\arabic{figure}}%
  \gdef\thetable{\@Alph\c@section.\arabic{table}}%
  \gdef\thesection{\appendixname\@Alph\c@section}%
  \@addtoreset{equation}{section}%
  \@addtoreset{table}{section}
  \@addtoreset{figure}{section}
}
\makeatother

\biboptions{sort&compress}

\begin{document}
\sloppy
\begin{frontmatter}
\title{Characteristics of $\langle a \rangle$  Screw Dislocations Basal Slip in Titanium}

\author{Ali Rida \corref{cor1}}
\ead{arida1@jhu.edu}
\author{Satish I. Rao \corref{}}
\author{Jaafar A. El-Awady \corref{cor1}}
\ead{jelawady@jhu.edu}

\cortext[cor1]{Corresponding authors}

\address{Department of Mechanical Engineering, Whiting School of Engineering, Johns Hopkins University, Baltimore, MD 21218, United States}

\begin{abstract}
Plasticity in hexagonal close packed (HCP) metals and alloys such as Titanium (Ti) and Zirconium (Zr) is carried out by the motion of $\langle a \rangle$ dislocations. Above room temperature, in situ transmission electron microscopy straining experiments have shown that basal slip of $\langle a \rangle$ screw dislocations in Ti is activated and its preponderance increases with temperature. To discern the mechanism of basal slip of these dislocations in pure Ti, molecular dynamics (MD) simulations and nudged elastic band (NEB) calculations were performed for a screw dislocation gliding on the basal plane. From MD simulations, basal and pyramidal slip were shown to be intertwined. The screw dislocation glides by kink pair nucleation and propagation mechanism on one of the two equiprobable pyramidal planes adjacent to the maximum resolved shear stress basal plane, before switching to the other pyramidal plane, thus leading to an average basal slip at low stress and/or high temperature. On the other hand, NEB calculations revealed that at low stresses kinks do not nucleate in the pyramidal plane but rather directly in the basal plane fully compatible with the motion of a screw dislocation confined to the basal plane, in agreement with experimental observations in Ti and Zr. The discrepancy between these two mechanisms is shown to be related to the thermal activation of the cross-slip from the pyramidal to the prismatic core configuration of the $\langle a \rangle$ screw dislocation.

\end{abstract}
\end{frontmatter}

\section{Introduction}
Secondary slip systems play an important role in the mechanical properties of metals and alloys with the hexagonal close packed (HCP) structure, where primary slip systems are not sufficient to achieve a complete three-dimensional deformation \cite{caillard2018glide}. At low temperature, plasticity in titanium (Ti) is controlled by the glide of $\langle a \rangle$ = 1/3$\langle 11 \bar{2} 0\rangle$ screw dislocations on the first order prismatic (prismatic-I)  $\lbrace \bar{1}100\rbrace$ planes \cite{akhtar1975prismatic, farenc1993situ, clouet2015dislocation}. In situ transmission electonic microscopy (TEM) straining experiments showed that the plastic deformation is governed by a jerky motion of long rectilinear screw dislocations on the prismatic-I planes  \cite{farenc1993situ,clouet2015dislocation}. This locking-unlocking thermally activated mechanism has been shown to be associated with the cross slip of screw dislocations from the prismatic-I to the first-order pyramidal (pyramidal-I) $\lbrace \bar{1}101 \rbrace$ plane and vice versa \cite{clouet2015dislocation}. The screw dislocation is sessile in its ground state pyramidal-I core configuration will need to cross slip to a higher energy metastable configuration dissociated on the prismatic-I plane where it can easily glide, then it will cross slip back to its sessile ground state core pyramidal-I core due to thermal activation (``locking-unlocking'' glide process).

Above room temperature, numerous transmission electron microscopy (TEM) observations showed evidence of secondary slip of $\langle a \rangle$ dislocations in the basal $(0001)$ planes and in the pyramidal-I planes \cite{shechtman1973orientation, akhtar1975basal, barkia2017situ, caillard2018glide}. Pyramidal-I slip was observed as a secondary slip when it is sufficiently stressed \cite{shechtman1973orientation,clouet2015dislocation}. The screw
dislocations were straight and glide slowly, which is a typical indication of high lattice friction on the pyramidal-I planes. Akhtar \cite{akhtar1975basal} observed the activation of basal slip above 500K in single crystals Ti,  basal slip was the only activated secondary slip system in addition to easy prismatic slip, with the relative ease of basal slip when increasing the temperature. In Ti-Al alloys with the atomic fraction of Al varying from $0.44 \%$ to $5.2 \%$, basal slip was observed at both temperature studied (298K and 500K)\cite{sakai1974basal}. The ratio of the basal to prismatic critical resolved shear stress was shown to decrease with increasing temperature and Al content. While the slip traces for prismatic-I and pyramidal-I slip were generally rectilinear on the surface even when the slip was located between them \cite{clouet2015dislocation, barkia2017situ}, the slip traces for basal slip were always wavy in Ti and Zr \cite{akhtar1975basal, barkia2017situ, caillard2018glide}. Caillard et al pointed out that basal slip is an elementary slip system in Ti and Zr \cite{caillard2018glide}. Their experimental observations showed that basal slip involves the slow and viscous motion of straight screw dislocations, in agreement with a kink-pair mechanism where screw dislocations are extended in the prismatic-I plane. Therefore, the waviness of the slip traces of basal slip is assumed experimentally to be composed of basal and prismatic-I slips. 

Besides experimental observations, molecular dynamics (MD) simulations of Maras and Clouet in pure Zr for  $\langle a \rangle$ screw dislocation gliding in the basal plane showed that basal slip is a combination of prismatic-I and pyramidal-I slip in the high stress regime typical for MD simulations \cite{maras2022secondary}. The gliding dislocation remains dissociated on the prismatic-I plane where it randomly glides without any friction stress and occasionally cross-slips out of its habit plane to one of the adjacent equiprobable pyramidal-I planes by kink pair mechanism in contrast with the above experimental observations. On the other hand, nudged elastic band (NEB) calculations revealed a change in the gliding mechanism at low stresses unreachable by MD simulations \cite{maras2022secondary}. At low stresses, kinks are no longer on the pyramidal-I plane but rather reside directly on the basal plane, fully compatible with the motion of a screw dislocation confined to the basal plane, as proposed by Caillard et al \cite{caillard2018glide}. Three dimensional dislocation dynamics simulations in Zr pillars also showed that the preponderance of plastic slip on basal and prismatic planes at different loading orientations and temperatures is mediated by a transitional dislocation, which is composed of glissile segments on  parallel prismatic-I planes connected by glissile super-jogs on basal planes \cite{li2023prismatic}. This is in agreement with the basal slip mechanism proposed by Caillard et al \cite{caillard2018glide}. 

The same configurations of $\langle a \rangle$ screw dislocation exist in Ti and Zr, except that their stability is opposite \cite{clouet2015dislocation}. In Zr, the core dissociated on the prismatic-I plane is the ground state configuration, whereas in Ti the ground state core is dissociated on the pyramidal-I plane. In addition, ab-initio calculations have shown that $\langle a \rangle$ screw dislocation dissociated in a basal plane is unstable in pure Ti \cite{kwasniak2019basal}, these calculations have shown that the mechanism for basal slip corresponds to a conservative motion of the stacking fault ribbon perpendicular to the dissociation plane. This mechanism is not compatible with the above experimental observations, where slip traces are fully resolved in the basal planes. Accordingly, this study focuses on characterizing basal slip activity in pure Ti.    


\section{Computational methods}
\subsection{Molecular dynamics simulations setup}
All MD simulations here were conducted using the open source large-scale Atomic/Molecular Massively Parallel Simulator (LAMMPS) \cite{thompson2022lammps}, with the Ti-Nb spline-like MEAM interatomic potential developed by Ehemann and Wilkins \cite{ehemann2017force}. This potential is referred to hereafter as the Ehemann potential. In an earlier study we showed that this potential predicts the correct core configurations for $\langle a \rangle$ screw dislocation in pure $\alpha$ Ti with the ground state core configuration dissociated on the pyramidal-I plane and a metastable core dissociated on the prismatic-I plane, in agreement with ab-initio calculations \cite{rida2022characteristics}. In addition, the Ehemann potential does not stabilize any core configuration for the $\langle a \rangle$ screw dislocation on the basal plane, also in agreement with ab-initio calculations \cite{rida2022characteristics}.

For the MD simulations in this study, a rectangular simulation cell with dimensions of $35 a_{lat} \times 40 \sqrt{3} a_{lat} \times 180 c$  (i.e., a total of 1,010,800 atoms) is constructed, where for pure Ti the lattice parameter on the basal plane is $a_{lat} = 2.936$ \AA, and the height parameter is $c = 1.578a_{lat}$. The $x-y$ plane of the simulation cell coincides with the crystal basal plane. Periodic boundary conditions were employed along both the simulation cell's $\textbf{x} = \left[ 1 1 \bar{2} 0 \right]$ and $\textbf{y} = \left[ \bar{1} 1 00 \right]$ axes, whereas free surfaces boundary conditions were imposed along the simulation cell $\textbf{z} = \left[ 0001\right]$ axis (i.e., the direction normal to the basal plane). A $\frac{1}{3}[11 \bar{2} 0]$ screw dislocation is then introduced at the center of the simulation cell using its anisotropic elasticity displacement field with its line direction along the periodic \textbf{x}-direction \cite{yoo1971numerical}. The elastic center of the dislocation is introduced between two narrowly spaced prismatic-I planes and halfway between two basal planes to reproduce the ground state pyramidal-I dissociation upon relaxation. A conjugate gradient minimization was then applied on the initial atomic positions to obtain the relaxed pyramidal-I core configuration. After that, the simulation cell is relaxed in the NVT ensemble at the desired temperature for 20 ps, followed by another relaxation in the NPT ensemble for 20 ps. 

A stress control loading is then applied on the simulation cell to drive the dislocation to glide on the basal plane. An applied shear stress, $\tau_{xz}$, is introduced on the simulation cell by applying forces in the \textbf{x}-direction on atoms in a thin layer having thickness of $\approx$ 10 \AA, at both boundaries in the \textbf{z}-direction. The forces at the top and bottom layers were equal and opposite, and the sum of the forces on either layer correspond to the applied pure shear stress. This shear stress leads to a Peach-Koehler force on the dislocation resolved in the \textbf{y}-direction in the basal plane, without any resolved shear stress in the prismatic plane to prevent glide on that plane. The shear stress is applied as a ramp shear stress at 5 MPa/ps until the desired shear stress is reached. The simulation cell is then relaxed at this applied stress for 1 ns, which is sufficiently long to allow the dislocation to reach its steady state velocity on the basal plane. All simulations are conducted with a 1 fs time step, applied stresses in the range 200-600 MPa, and for temperatures in the range 400-600 K. 


\subsection{Nudged elastic band calculations of the Peierls barrier and $\langle a \rangle$ screw dislocation core configuration}
The Peierls barrier for the slip on the prismatic-I, pyramidal-I, and basal planes for an $\langle a \rangle$ screw dislocation is calculated using Nudged elastic band (NEB) calculations \cite{henkelman2000improved, henkelman2000climbing, maras2016global}. For these calculations a rectangular simulation cell with dimensions of $5 a_{lat} \times 40 \sqrt{3} a_{lat} \times 45 c$  (i.e., a total of 36,630 atoms) is constructed. Periodic boundary conditions are employed along the $\textbf{x} = \left[ 1 1 \bar{2} 0 \right]$ direction (i.e., the dislocation line direction), whereas free surfaces boundary conditions are imposed along the  $\textbf{y} = \left[ \bar{1} 1 00 \right]$ direction, and along the normal to the basal plane (i.e., the $\textbf{z} = \left[ 0001\right]$ direction). 

\begin{figure}
    \centering
    \includegraphics[width= 0.6\linewidth]{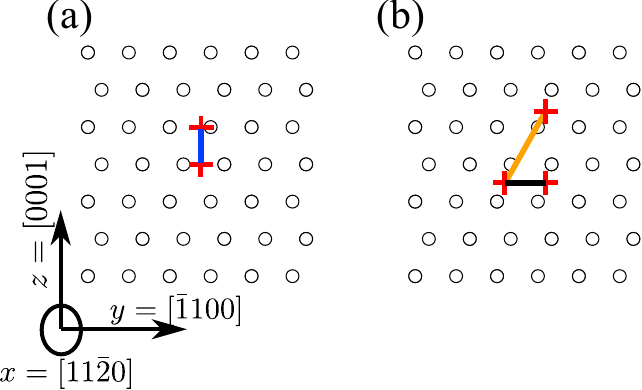}
    \caption{The atomic positions on the $(11 \bar{2}0)$ plane for pure Ti where the red crosses indicate the position of the center of the $\langle a \rangle$ screw dislocation in the initial and final Peierls valley necessary for slip barrier calculation using NEB (a) for prismatic-I slip and (b) for basal and pyramidal-I slip. The blue, black, and orange solid lines show the initial gliding path used in NEB calculations between the position of the centers of the dislocation in the initial and final replicas for prismatic-I, basal, and pyramidal-I slip, respectively.}
    \label{Fig:NEB_trajectory}
\end{figure}

For prismatic-I slip barrier calculation, the centers of the $\langle a \rangle$ screw dislocation in the initial and final configurations are located in the same prismatic-I plane as shown in Figure \ref{Fig:NEB_trajectory}(a). On the other hand, in the case of pyramidal-I slip barrier calculations, both centers are located on the narrowly spaced pyramidal-I plane, as shown in Figure \ref{Fig:NEB_trajectory}(b) by the orange line connecting both centers. Finally, for basal slip barrier calculation both centers are located on the same basal plane, as shown in Figure \ref{Fig:NEB_trajectory}(b) by the black line connecting both centers. Intermediate replicas are built by linearly interpolating the atomic coordinates between the initial and final state. Different numbers of intermediate replicas where chosen in each case for faster convergence. In this study, the NEB calculations are carried out with a spring constant, K = 0.1 eV/$\AA^2$, connecting the replicas, until the maximum value of the norm of the 3N-dimensional force vector of each image has dropped below $10^{-3}$ eV/$\AA$. The reaction coordinate $\zeta$ for each replica is defined as:
\begin{equation}
    \zeta = \frac{\left(\Vec{X} - \Vec{X}^I \right).\left(\Vec{X}^{F} - \Vec{X}^I \right)}{\| \Vec{X}^{F} - \Vec{X}^I  \|^2}
\end{equation}
where $\Vec{X}$, $\Vec{X}^I$, and $\Vec{X}^{F}$ are the 3N vectors defining atomic positions for the intermediate, initial, and final replicas, respectively.

\subsection{Nudged elastic band calculations of the enthalpy barrier for kink nucleation during basal slip}
The variation of the enthalpy barrier for kink nucleation with the applied stress during basal slip is computed using the NEB method \cite{henkelman2000improved, henkelman2000climbing, maras2016global}. For these calculations, the initial simulation cell of the previous NEB calculations was sufficiently extended in the \textbf{x}-direction to $56 a_{lat}$ to capture kink pair nucleation. Periodic boundary conditions were employed along the $\textbf{x} = \left[ 1 1 \bar{2} 0 \right]$ direction, whereas free surfaces boundary conditions were imposed along the  $\textbf{y} = \left[ \bar{1} 1 00 \right]$ and $\textbf{z} = \left[ 0001\right]$ directions. 

In these calculations 24 replicas of the dislocation along the gliding path on the basal plane between the initial and final Peierls valleys were designed. Unlike the previous NEB calculations, the 22 intermediate replicas of the dislocation are not linearly interpolated between the initial and final states. Instead, the dislocation structure in each replica contains kinks with segments lying in both Peierls valleys, with the length of the dislocation segment in the final valley being proportional to the replica index along the path. In this way, the dislocation will not simply move as a whole like in the previous NEB calculations, but instead in a kink-pair nucleation and propagation mechanism. These calculations are also performed using a spring constant connecting the replicas of K = 0.1 eV/$\AA^2$, in addition a perpendicular spring with a spring constant K = 0.5 eV/$\AA^2$ is used to prevent the path from becoming too strongly kinked and thereby improve convergence especially under applied stress \cite{maras2016global}. These NEB calculations are conducted until the maximum value of the norm of the 3N-dimensional force vector of each image drops below $10^{-2}$ eV/\AA.

Applying shear stress will shift the enthalpy barrier due to the work of the Peach-Koehler stress. The same shear stress procedure as in the above MD case was used to load the simulation cell for these NEB calculations. However, here the simulation cell was directly loaded to the desired shear stress while performing NEB energy minimization.

\section{Results}
\subsection{The dislocation mobility}
The dislocation position was determined using the adaptive common neighbor analysis algorithm (ACNA) implemented in OVITO \cite{stukowski2012structure}. Atoms belonging to the core of the dislocation are selected and the dislocation position is determined as the average of the coordinates of these atoms in the $ (11\bar{2}0)$ plane. The dislocation position is also extracted from these simulations using the dislocation extraction algorithm implemented in OVITO, which gives identical position \cite{stukowski2012automated}.



\begin{figure}
    \centering
    \includegraphics[width= 0.9\linewidth]{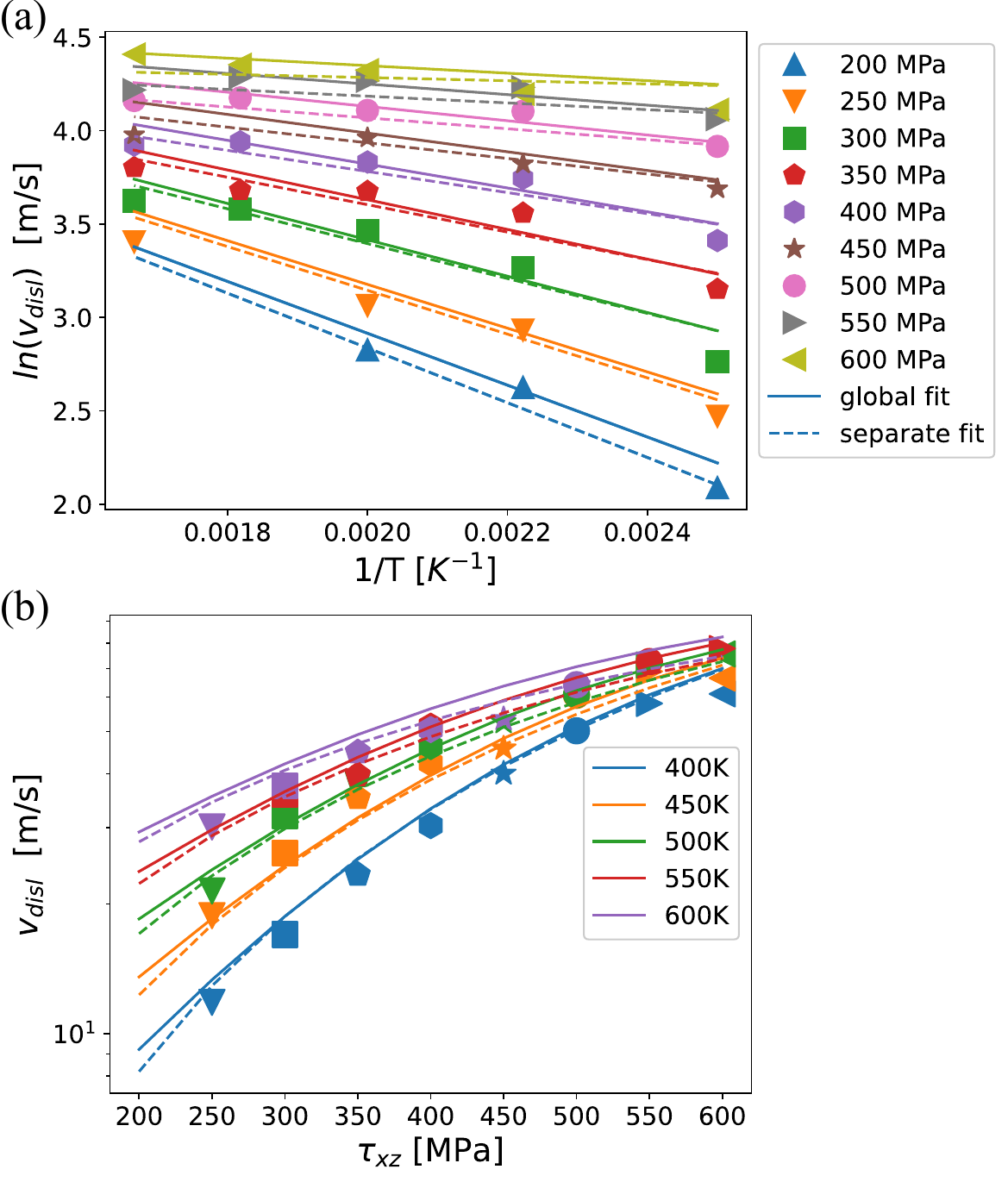}
    \caption{Dislocation velocities extracted versus: (a) applied shear stress and (b) temperature. The symbols represent the MD simulation predictions. The dashed lines in (a) are the fits to an Arrhenius equation to extract the activation enthalpy and velocity prefactor for each applied shear stress as shown in  Eq. (\ref{eq:thermal_activation}). The dashed lines in (b) are the plots of Eq. (\ref{eq:analytical_velocity_vs_T_tau}) using the separate fit extracted parameters shown in table \ref{table:fitting_results}. The solid lines in (a) and (b) are the global fits to Eq. (\ref{eq:analytical_velocity_vs_T_tau}) to all the dislocation velocities at different shear stress and temperature.}
    \label{Fig:disl_velocity}
\end{figure}

\begin{figure}
	\centering
	\includegraphics[width= 0.7\linewidth]{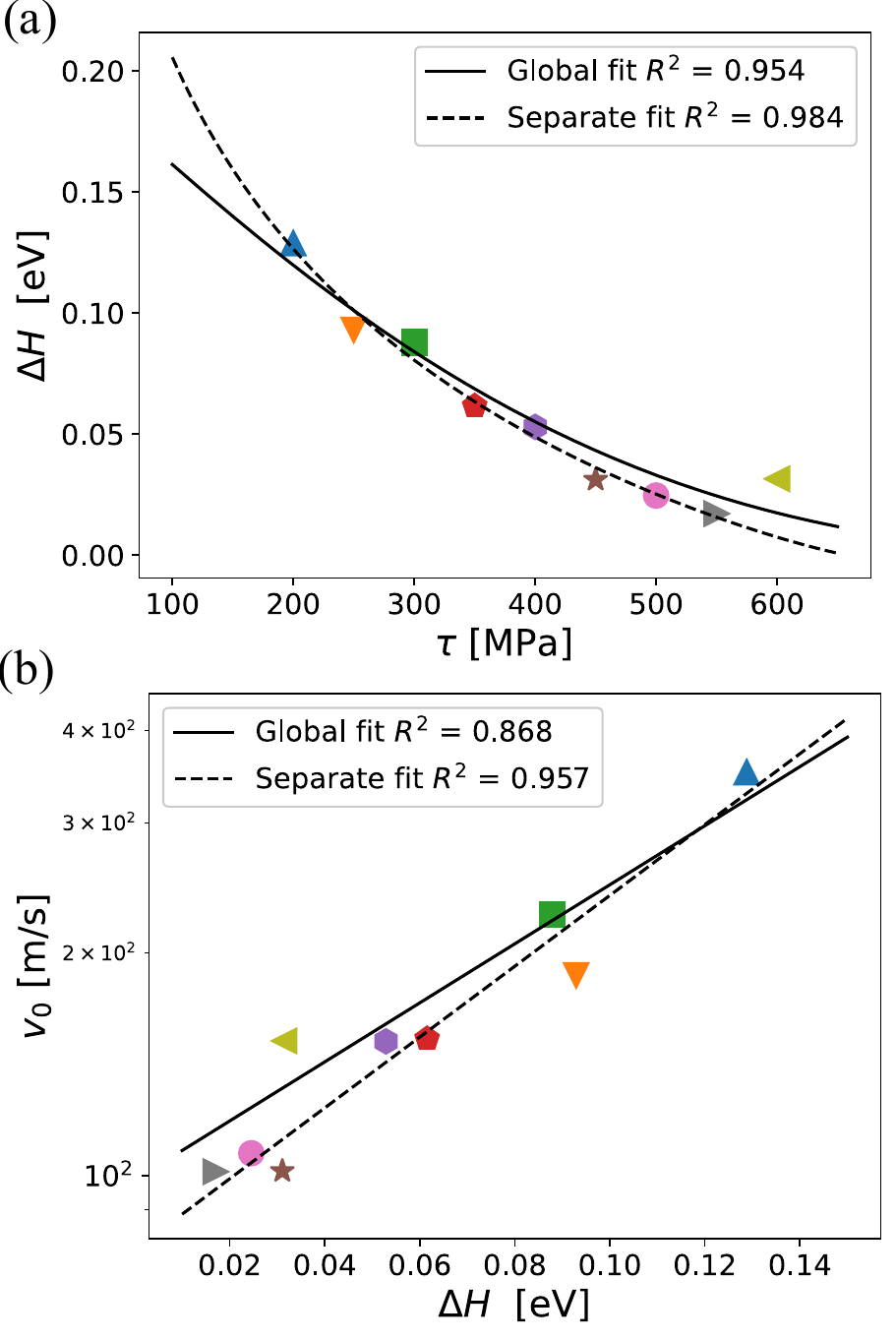}
	\caption{(a) Activation enthalpy as function of the applied shear stress; and (b) velocity prefactor as function of the activation enthalpy. These thermal activation law parameters were computed from the Arrhenius equation fits in Figure \ref{Fig:disl_velocity}(a). The dashed lines in (a) and (b) are the fit of Eq. (\ref{eq:Kocks_law}) and Eq. (\ref{eq:velocity_prefactor}) to the activation enthalpy and velocity prefactor results, respectively. The solid lines in (a) and (b) are obtained through the global fit of Eq. (\ref{eq:analytical_velocity_vs_T_tau}) to all MD results in Figure \ref{Fig:disl_velocity}.}
	\label{Fig:enthalpy_v0_vs_tau}
\end{figure}

Dislocation velocities deduced from the MD simulations as function of stress and temperature in the studied ranges are shown by symbols in Figure \ref{Fig:disl_velocity}. Figure \ref{Fig:disl_velocity}(a) shows an Arrhenius plot of the dislocation velocities for each applied shear stress at different temperatures, whereas Figure \ref{Fig:disl_velocity}(b) shows the variation of the dislocation velocities with the applied shear stress at a constant temperature. In these simulations, the dislocation glide by a thermal activation kink pair mechanism. From the variation of the velocity for a given applied shear stress with temperature in Figure \ref{Fig:disl_velocity} (a) the activation enthalpy can be deduced as \cite{caillard2003thermally}:
\begin{equation}
    V_{disl}(\tau, T) = V_0 \exp \left\{-\Delta H(\tau)/ K_B T \right \}
    \label{eq:thermal_activation}
\end{equation}
where $V_0$ is a velocity prefactor, $K_B$ the Boltzman constant, $\Delta H$ the activation enthalpy depending on the shear stress $\tau$, and $T$ is temperature. The fit of Eq. (\ref{eq:thermal_activation}) to MD dislocation velocities is shown in dashed lines in Figure \ref{Fig:disl_velocity}(a). The resultant activation enthalpy and velocity prefactor for each applied shear stress are shown by symbols in Figure \ref{Fig:enthalpy_v0_vs_tau}. 

The variation in the activation enthalpy versus the shear stress can be described by a Kocks law \cite{kocks1975progress}:
\begin{equation}
    \Delta H(\tau) = \Delta H_0 \left\{  1 - \left( \frac{\tau}{\tau_p}\right)^p \right\}^q
    \label{eq:Kocks_law}
\end{equation}
where $\Delta H_0$ and $\tau_p$ are the activation enthalpy for zero applied shear stress and the Peierls stress, respectively. Additionally, $0 \leq p \leq 1 $ and $1 \leq q \leq 2$ are fitting constants. The fit of Eq. (\ref{eq:Kocks_law}) to the variation of the activation enthalpy with the applied shear stress is shown by dashed line in Figure \ref{Fig:enthalpy_v0_vs_tau}(a). The extracted parameters are summarized in Table \ref{table:fitting_results} in the separate fit column. The predicted parameters are very sensitive to the way the fitting is performed as will be discussed below. Therefore, they should only be regarded in the studied range of stress and temperature and are inaccurate for any extrapolation beyond this range. Similar to the pure Zr case \cite{maras2022secondary}, the logarithm of the velocity prefactor display a linear relation with the activation enthalpy, as shown in Figure \ref{Fig:enthalpy_v0_vs_tau}(b). Following Maras and Clouet \cite{maras2022secondary}, the velocity prefactor can be written as a product of a length dependent constant term multiplied by a stress dependent entropy term as:
\begin{equation}
    V_0 = \frac{L_D}{b} \nu \lambda_p \exp \left( \frac{\Delta S(\tau)}{K_B}\right)
\end{equation}
where $L_D$ is the dislocation line length, $\lambda_p = a \sqrt{3}/2$ is the distance between two Peierls valleys in the gliding direction, $\nu$ is the attempt frequency, and $\Delta S$ the activation entropy. Assuming that the activation entropy is proportional to the activation enthalpy $\Delta S = \Delta H/T_m$, the Meyer-Neldel compensation rule will thus lead to \cite{maras2022secondary}:
\begin{equation}
    \log[V_0(\tau)] = \log \left( \frac{L_D \nu \lambda_p}{b} \right) + \frac{\Delta H(\tau)}{K_B T_m}
    \label{eq:velocity_prefactor}
\end{equation}
The fit of Eq. (\ref{eq:velocity_prefactor}) to the variation of the velocity prefactor with the activation enthalpy is shown in dashed line in \ref{Fig:enthalpy_v0_vs_tau}(b). The resulting parameters from this fitting procedure are $T_m$ and $\nu$ which are shown in the separate fit column in Table \ref{table:fitting_results}. The attempt frequency, and the activation enthalpy at zero applied stress are lower than the pure Zr case \cite{maras2022secondary}. This is related to the lower activation enthalpy versus stress reported here in comparison to the pure Zr case \cite{maras2022secondary}, even though the applied shear stress range for Ti is smaller than that for Zn. Therefore, one can expect that the basal slip mechanism is different in Ti and Zr from these dynamical simulations as will be discussed in Section \ref{Sec:discussion}. 
Plugging Eqs. (\ref{eq:Kocks_law}) and (\ref{eq:velocity_prefactor}) into Eq. (\ref{eq:thermal_activation}) leads to:
\begin{equation}
    V_{disl}(\tau, T) = \frac{L_D}{b}\nu \lambda_p \exp \left\{ - \frac{\Delta H_0}{K_B T} \left[1-  \left( \frac{\tau}{\tau_P}\right)^p\right]^q \left[ 1 - \frac{T}{T_m}\right] \right\}
    \label{eq:analytical_velocity_vs_T_tau}
\end{equation}
The fitting of the MD dislocation velocities by this analytical expression using the parameters of the separate fit column in Table \ref{table:fitting_results} is shown by the dashed lines in Figure \ref{Fig:disl_velocity}(b). A reasonable agreement is obtained with the MD results. To assess the validity of this separate fitting procedure, Eq. (\ref{eq:analytical_velocity_vs_T_tau}) is directly fitted to the MD dislocation velocities where the four parameters of the Kocks law and the two parameters of the Meyer-Neldel rule are considered variables (global fit). The resulting fitting parameters are shown in the global fit column in Table \ref{table:fitting_results}. This fitting is shown by solid lines in Figure \ref{Fig:disl_velocity} and \ref{Fig:enthalpy_v0_vs_tau}. It is clear that this fitting procedure is not better than the previous procedure as shown in Figure \ref{Fig:disl_velocity}(b) and by the $R^2$ values shown in Figure \ref{Fig:enthalpy_v0_vs_tau}. The activation enthalpy at 0 applied shear stress, $\Delta H_0$, and the Peierls stress, $\tau_p$, varies significantly between both fitting procedures. For this reason, these thermal activation parameters should only be considered for the stress and temperature regime studied here and should not be extrapolated to lower or higher applied shear stresses and temperatures.

\begin{table}
    \centering
    \caption{Parameters of the Kocks law and the Meyer-Neldel compensation rule deduced from the fitting of the MD extracted results in Figure \ref{Fig:enthalpy_v0_vs_tau} using Eqs. (\ref{eq:Kocks_law}), and (\ref{eq:velocity_prefactor}), or by a global fit of the MD results in Figure \ref{Fig:disl_velocity} using Eq. (\ref{eq:analytical_velocity_vs_T_tau}).  }
    \begin{tabular}{c c c}
        \hline
                & separate fit & global fit \\
        \hline                  
        $\Delta H_0$ [eV] & 1.8  & 0.2   \\
        $\tau_p$ [MPa]  & 658 &  980 \\
        p & 0.0845 & 1.114 \\
        q & 1.13 & 2.85 \\
        $\nu$ [THz] & 0.00892 &  0.0111 \\
        $T_m$ [K] & 1051  & 1261 \\
	\hline 		   
    \end{tabular}
    \label{table:fitting_results}
\end{table}

\subsection{The predicted dislocation trajectory}
\begin{figure}
    \centering
    \includegraphics[width= \linewidth]{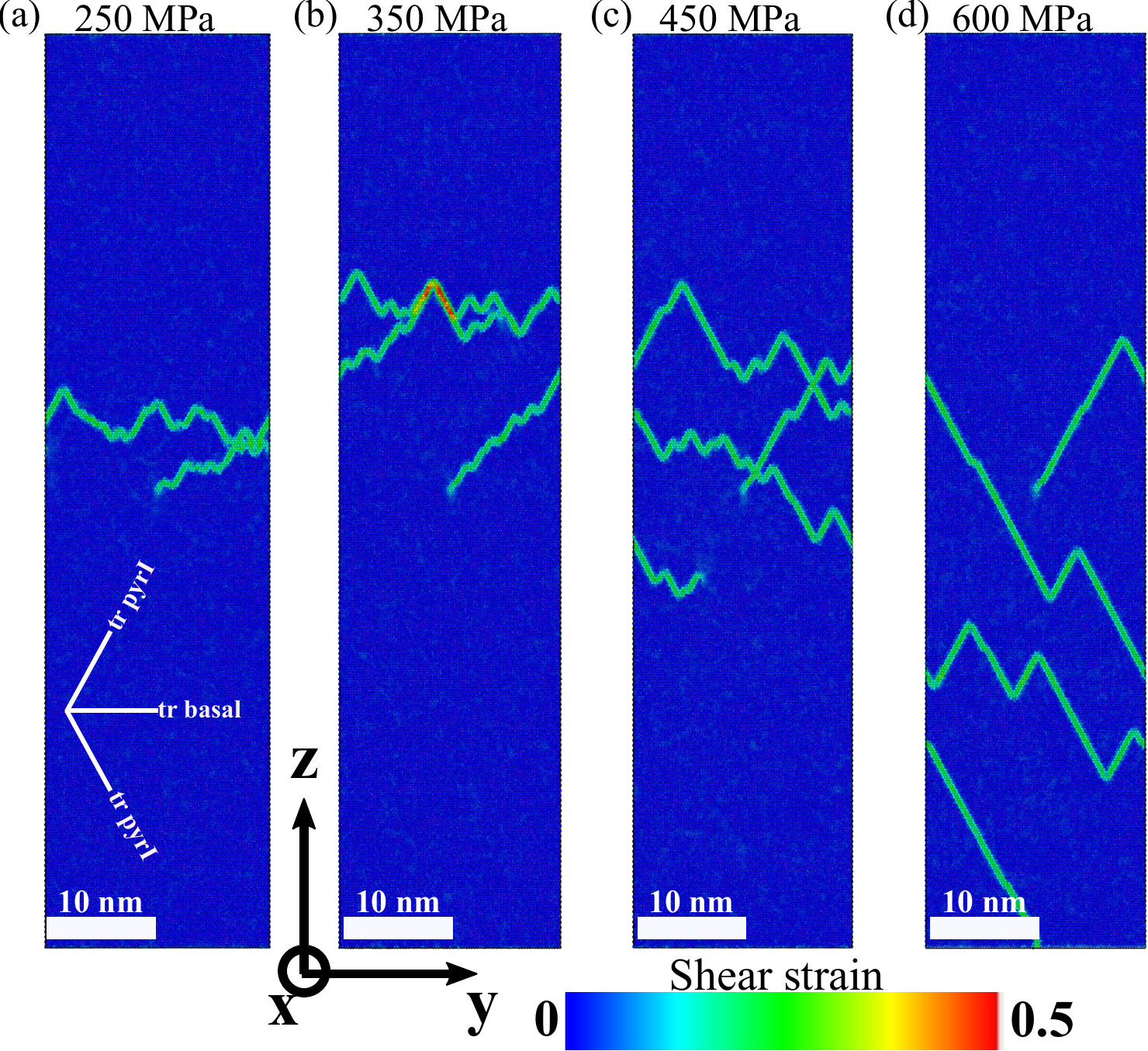}
    \caption{Side view of the atomic configurations after 1 ns relaxation at 600K and at an applied shear stress of: (a) 250 MPa; (b) 350 MPa; (c) 450 MPa; and (d) 600 MPa. The atoms are colored by the atomic shear strain modifier in OVITO with respect to the relaxed atomic configuration at 600K and zero applied shear stress. The trajectory of the dislocation is clearly visible by the green connected atoms in these snapshots. It should be noted that the dislocation makes several passes through the periodic boundary of the simulation cell. The traces of the basal, and the two pyramidal-I planes are also shown by the lines in (a).}
    \label{Fig:trajectory_vs_stress}
\end{figure}

The atomic shear strain modifier in OVITO is a good indicator to monitor the swept area and subsequently the trajectory of the dislocation \cite{shimizu2007theory}. The atomic configurations after 1 ns relaxation at different applied shear stresses and temperatures were extracted and colored by the atomic shear strain with respect to the reference configuration at the same temperature but without applied shear stress. In this way the trajectory of the dislocation can be clearly highlighted by the atoms having large shear strain indicating that these atoms were swept by the dislocation. The effect of the applied stress on the trajectory of the dislocation at 600K is shown in Figure \ref{Fig:trajectory_vs_stress}.  In these simulations, the only non-null component of the applied shear stress is $\tau_{xz}$, hence the two pyramidal-I planes see exactly the same resolved shear stress, leading to an equal probability for dislocation glide on either of these two planes. The dislocation glide is mitigated by the kink pair mechanism on the two adjacent equi-probable pyramidal-I planes. However, at small applied shear stresses (e.g, 250 MPa) the dislocation is transitioning from one pyramidal-I plane to another with small gliding steps on each plane, leading to an approximate average basal slip. On the other hand, at high stresses (e.g, 600 MPa) the dislocation is gliding on each pyramidal-I plane for a longer period of time before transferring to the other pyramidal-I plane deviating from an averagely basal slip. It should be noted that at the transfer from one pyramidal-I plane to another, the dislocation is adjusting its atomic configuration to the prismatic metastable core configuration \cite{rida2022characteristics}.

\begin{figure}
    \centering
    \includegraphics[width= \linewidth]{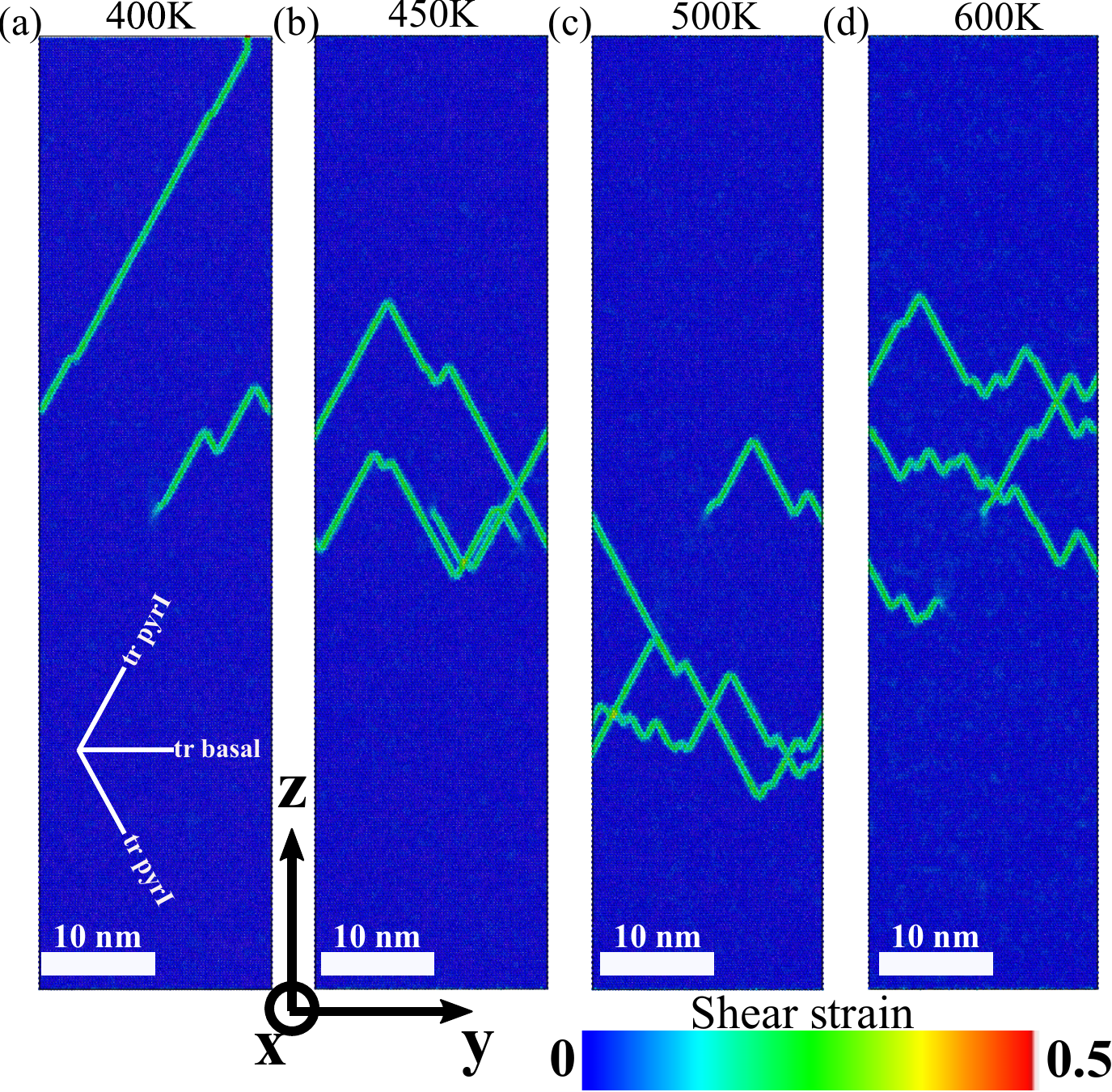}
    \caption{Side view of the atomic configurations after 1 ns relaxation at a constant shear stress of 450 MPa and: (a) 400K; (b) 450K; (c) 500K; and (d) 600K. The atoms are colored by the atomic shear strain modifier in OVITO with respect to the relaxed atomic configuration at each temperature and zero applied shear stress. The trajectory of the dislocation glide is clearly visible by the green connected atoms on these snapshots. It should be noted that the dislocation makes several passes through the periodic boundary of the simulation cell. The traces of the basal, and the two pyramidal-I planes are also shown by the lines in (a).}
	\label{Fig:trajectory_vs_temp}
\end{figure}

Figure \ref{Fig:trajectory_vs_temp} shows the effect of temperature on the trajectory of the dislocation for the same applied shear stress of 450 MPa. The dislocation is shown to always glide by transitioning between the two pyramidal-I planes adjacent to the basal planes, which has the maximum resolved shear stress, irrespective of the temperature. However, at high temperatures (e.g, 600k) the dislocation glides for a short distance on each pyramidal-I plane, resulting in an approximate average basal slip response. On the other hand, at low temperatures (e.g, 400K) the dislocation glides for longer distances on the pyramidal-I planes. Therefore, an equivalence between the stress and temperature dependency can be concluded, where the high temperature regime for dislocation motion is equivalent with the low stress regime and the low temperature regime is equivalent to the high stress regime.


\subsection{Peierls barrier for $\langle a \rangle$ screw dislocation slip on prismatic-I, pyramidal-I, and basal planes}
\begin{figure}
    \centering
    \includegraphics[width= \linewidth]{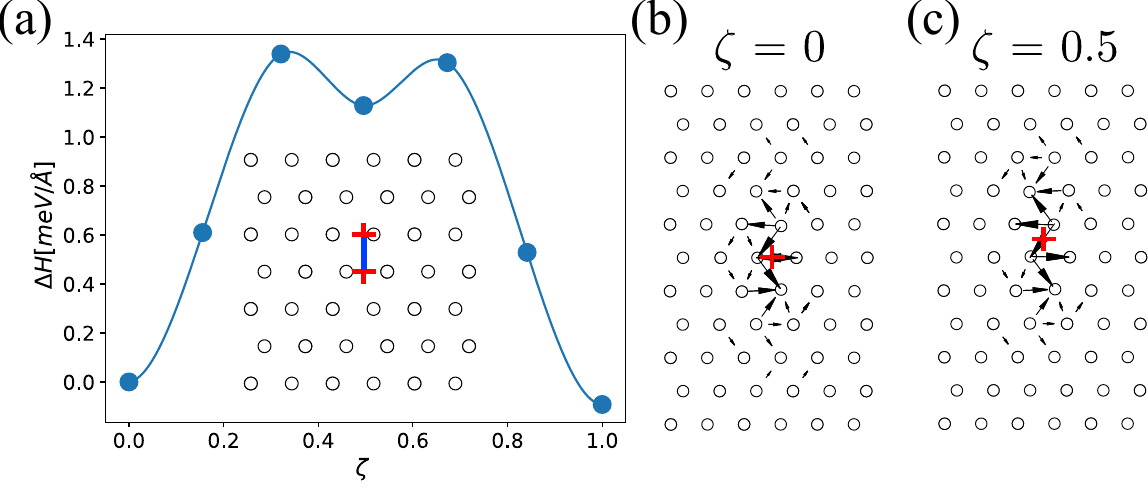}
    \caption{(a) The energy barrier for the glide of a prismatic $\langle a \rangle$ screw dislocation on the prismatic-I plane. The inset shows the atomic positions on the $(11\bar{2}0)$ plane where the two red crosses indicate the position of the center of the dislocation in the initial and final Peierls valley. The blue line shows the initial gliding path between the position of the centers of the dislocation in the initial and final replicas. The differential displacement plots of the (b) stable initial and (c) the meta-stable core configurations are also shown. The same minimum energy path was found after convergence of the NEB calculation.}
    \label{Fig:prismatic_slip_barrier}
\end{figure}

Figure \ref{Fig:prismatic_slip_barrier}(a) shows the energy barrier encountered by a prismatic $\langle a \rangle$ screw dislocation gliding on the prismatic-I plane as predicted from NEB calculations. The energy barrier is observed to be low (i.e., 1.3 meV/$\AA$). Additionally, a meta-stable core configuration is observed to form halfway along the minimum energy path (MEP) on the glide plane. Differential displacement plots of the stable initial and the meta-stable core configurations are also shown in Figure \ref{Fig:prismatic_slip_barrier}(b) and (c). It is observed that while the center of the dislocation is on the basal plane between two widely spaced prismatic-I planes in the stable initial core configuration (see Figure \ref{Fig:prismatic_slip_barrier}(b)), the center of the dislocation for the meta-stable configuration lies halfway between two basal planes and between two widely spaced prismatic-I planes (see Figure \ref{Fig:prismatic_slip_barrier}(c)). A very small energy difference is observed between both cores (i.e., 1.13 meV/$\AA$). This indicates that the friction stress on the prismatic-I plane is very small, which is in agreement with the experimental observation that prismatic slip is the primary deformation mode at low temperatures \cite{naka1988low, farenc1993situ, clouet2015dislocation}. These two cores are also in excellent agreement with the prismatic cores found in ab-initio calculations \cite{clouet2015dislocation}, however, the energy ordering of both cores is opposite with respect to the current NEB calculations. The very small energy difference between both cores in both the ab-initio calculations (i.e., 0.4 meV/$\AA$ \cite{clouet2015dislocation}) and the current calculations indicates a high mobility of $\langle a \rangle$ screw dislocation on the prismatic-I planes, with the core structure constantly switching between these two cores.     

\begin{figure}
    \centering
    \includegraphics[width= 0.7\linewidth]{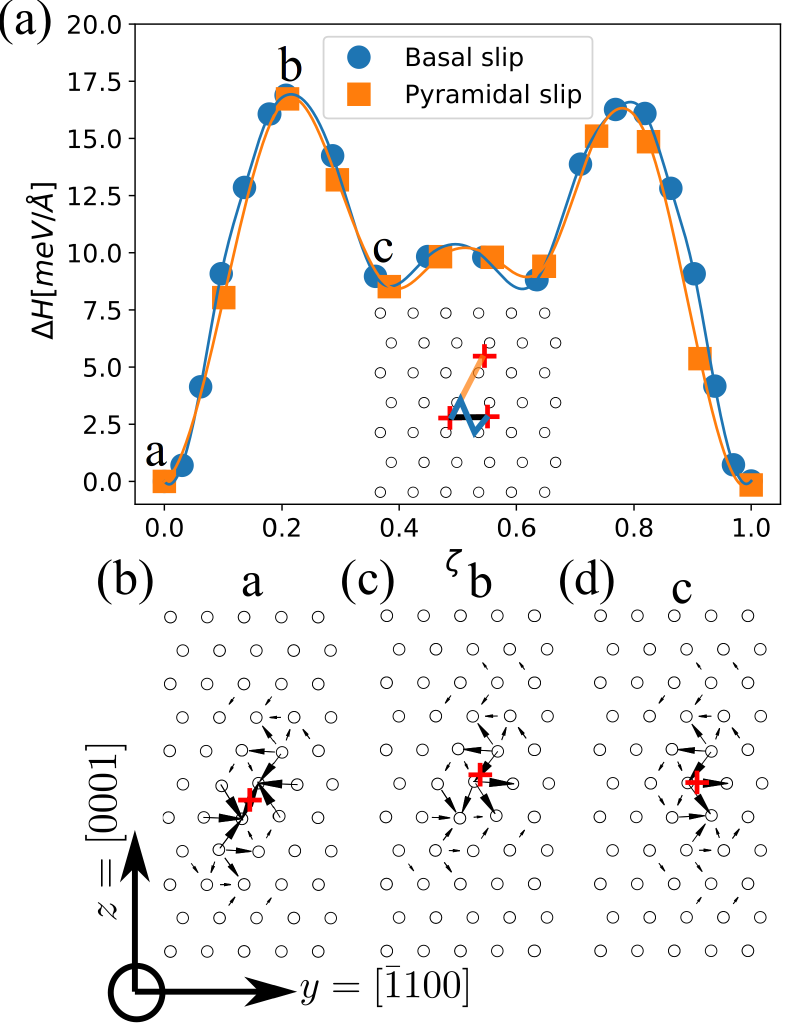}
    \caption{(a) The energy barriers encountered by pyramidal-I $\langle a \rangle$ screw dislocations when gliding on the basal or pyramidal-I planes. The inset shows the atomic positions on the $(11\bar{2}0)$ plane where the three red crosses indicate the position of the center of the dislocation in the initial and final Peierls valley on the pyramidal-I and basal planes. The  black and orange lines in the inset of the Figure show the initial gliding path between the position of the centers of the dislocation in the initial and final replicas on the basal and pyramidal-I plane, respectively. The Differential displacement plots of the (b) most stable initial, (c) the unstable, and (d) the meta-stable core configurations are also shown. The blue line shows the converged minimum energy path after the convergence of the NEB calculation for basal slip. The initial path for slip on the pyramidal-I plane stayed the same after NEB calculation's convergence.}
    \label{Fig:basal_pyramidal_slip_barrier}
\end{figure}

The energy barrier for the slip of the $\langle a \rangle$ screw dislocation on the pyramidal-I and basal planes are shown in Figure \ref{Fig:basal_pyramidal_slip_barrier}(a). The energy barrier is found to be similar for both slip systems. While the MEP for the glide on the pyramidal-I plane after convergence of the NEB calculation is the same as before performing NEB minimization (orange line in Figure \ref{Fig:basal_pyramidal_slip_barrier}), the converged MEP for basal slip shows a zigzag shape onto the two adjacent pyramidal-I planes (blue line in the insert of Figure \ref{Fig:basal_pyramidal_slip_barrier}(a)). This clearly confirms that slip on the pyramidal-I and basal slip systems are intertwined. 

The differential displacement plots of the stable, unstable, and meta-stable core configurations along the MEP are shown in Figures \ref{Fig:basal_pyramidal_slip_barrier}(b)-(d), respectively. The excess energy of the meta-stable prismatic-I core configuration with respect to the stable pyramidal-I core is $\approx$ 8.5 meV/$\AA$, which is slightly higher than the previous value reported from MS calculations of $\approx$ 6.4 meV/$\AA$ \cite{rida2022characteristics}, and in ab-initio calculations of 5.7 meV/$\AA$ \cite{clouet2015dislocation}. However, this value is closer to ab-initio calculations of Tsuru et al, which was $\approx$ 7.8 meV/$\AA$ \cite{tsuru2022dislocation}. On the other hand, the unstable core configuration also shows a pyramidal-I spreading where the dislocation center is very close to an atomic position on the pyramidal-I plane, which is between two widely spaced prismatic-I planes, in contrast with the ground state core configuration predicted by this potential. Interestingly, this core configuration is the most stable core found in ab-initio calculations \cite{clouet2015dislocation}. Therefore, the most stable and unstable pyramidal-I core configurations given by the Ehemann potential are opposite to DFT calculations \cite{clouet2015dislocation}. To the authors knowledge, none of the interatomic potential in the literature were shown to reproduce the ground state core configuration found in ab-initio calculations at zero applied stress with the dislocation center between the two widely spaced prismatic-I planes halfway between two basal planes including the recently developed Deep Potential for Ti \cite{wen2023modelling} which reproduces similar pyramidal-I core configurations to the Ehemann potential. 

Finally, Unlike prismatic-I slip, the energy barrier for pyramidal-I or basal slip is high ($\approx$ 17 meV/$\AA$), which indicates a high lattice friction on these planes, in agreement with experimental observations showing straight $\langle a \rangle$ dislocations aligned toward their screw character on the pyramidal-I planes \cite{farenc1993situ, clouet2015dislocation}. This energy barrier is also in reasonable agreement with the barriers calculated from ab-initio calculations for pyramidal-I and basal slips, which are in the range of 11-18 meV/$\AA$ \cite{clouet2015dislocation, kwasniak2019basal, tsuru2022dislocation}. 
    
\subsection{Enthlapy barrier for kink nucleation}

NEB calculations were performed here to study the mechanisms controlling basal slip of $\langle a \rangle$ screw dislocations at different applied shear stresses to extend the studied stress range beyond the those reached by MD simulations.  The goal is to determine the MEP of the screw dislocation while it glides from one Peierls valley to another on the basal plane. This will enable determining the effect of the applied stress on the energy barrier for kink-pair nucleation, as well as describe the variation in the shape of the dislocation on the MEP for basal slip.

\begin{figure}
    \centering
    \includegraphics[width= 0.6\linewidth]{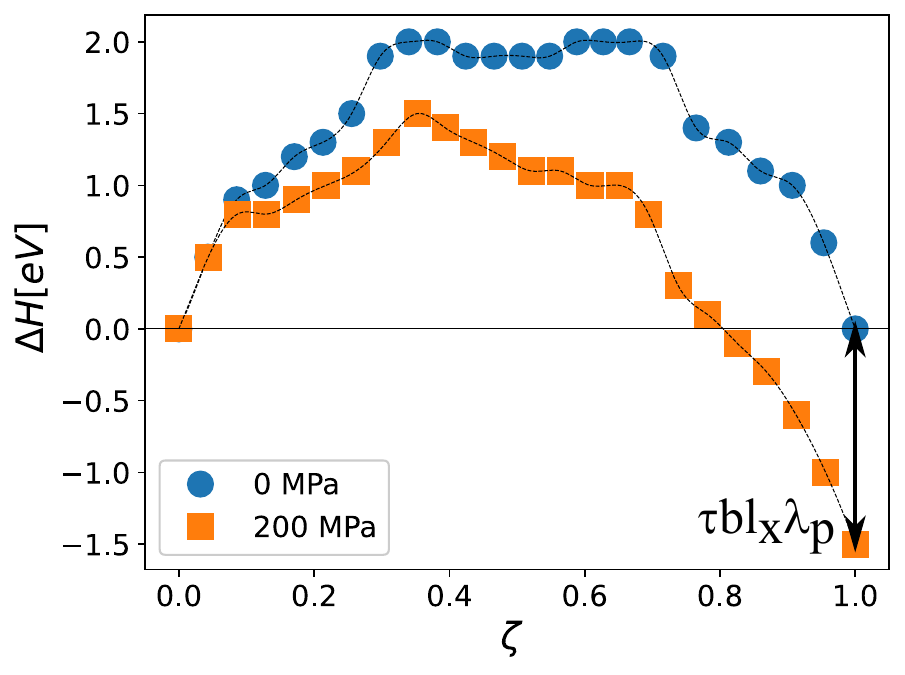}
    \caption{The enthalpy barrier $\Delta H(\zeta)$ as predicted from  NEB calculations versus the normalized NEB reaction coordinate, $\zeta$, for basal $\langle a \rangle$ screw dislocation glide along the MEP between the two Peierls valley at 0, and 200 MPa applied shear stress.    
    }
    \label{Fig:enthalpy_barrier_vs_tau}
\end{figure}

Figure \ref{Fig:enthalpy_barrier_vs_tau} shows the variation of the enthalpy barrier versus the normalized NEB reaction coordinate, $\zeta$, along the MEP between the two Peierls valley for 0, and 200 MPa applied shear stresses. The highest enthalpy corresponds to the barrier of nucleating a kink pair on the basal plane. For the case with zero applied shear stress, this barrier is 2.0 eV with the enthalpy barrier exhibit a constant plateau at this value with small undulation corresponding to the migration of kinks along the dislocation line with a small migration energy of 0.1 eV. The migration of kinks is thus much easier than their formation. 

When a constant shear stress of 200 MPa is applied, the work of the Peach-Koehler force will shift the enthalpy barrier by:
\begin{equation}
    \Delta H (\zeta) = \Delta U(\zeta) - \tau b l_x \lambda_p \zeta
\end{equation}
where $\Delta U(\zeta)$ is the internal energy, and $l_x$ is the dislocation line length. The second term on the right hand side of the equation represents the work done. The total work done when the dislocation glides a total distance $\lambda_p$ to the final Peierls valley is thus, $\tau b l_x \lambda_p$, as shown in Figure \ref{Fig:enthalpy_barrier_vs_tau}. 

\begin{figure}
    \centering
    \includegraphics[width= 0.8 \linewidth]{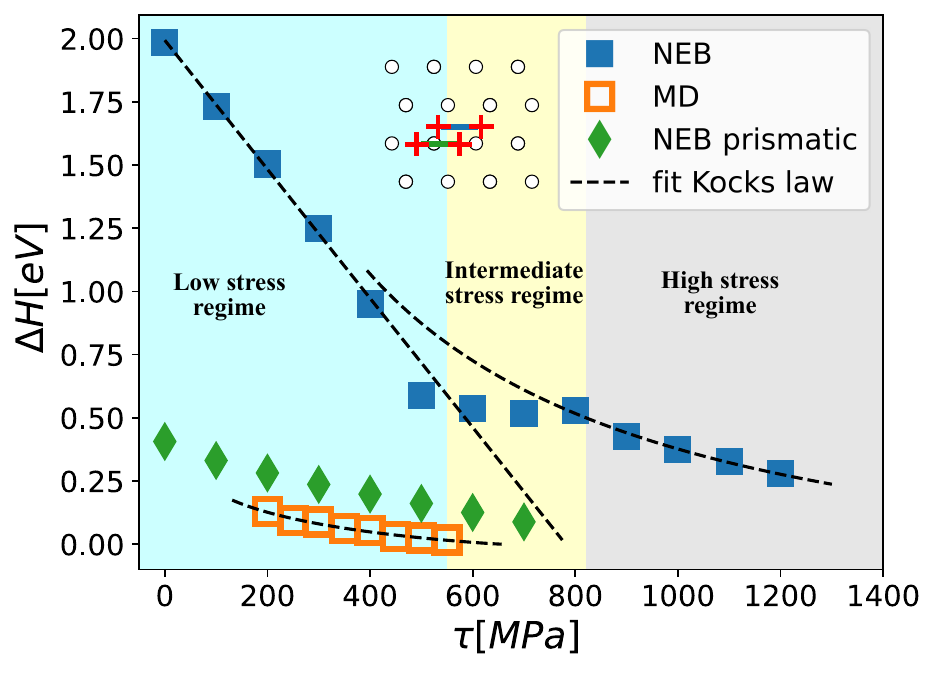}
    \caption{The activation enthalpy, $\Delta H$, versus the applied shear stress, $\tau$, for the basal slip of $\langle a \rangle$ screw dislocation. The NEB predictions of the pyramidal-I (blue squares) and the prismatic-I (green diamonds) core configurations are shown along with the MD predictions (orange empty squares). The dashed lines represent the Kocks law with different fits for the different stress regimes. The inset shows the atomic positions on the $(11\bar{2}0)$ plane where the blue and green lines highlight the initial gliding path between the position of the centers of the $\langle a \rangle$ screw dislocation in the initial and final NEB replicas on the basal plane. The shaded areas separate the three regimes where a change in the basal slip mechanism of $\langle a \rangle$ screw dislocation is happening. The reader is referred to the text for more detail.}
    \label{Fig:Delta_H_vs_tau}
\end{figure}

The variation of the enthalpy barrier with the applied shear stress is shown in Figure \ref{Fig:Delta_H_vs_tau}. Two transitions in the activation enthalpy can be observed around 550 and 800 MPa. Below 550 MPa the activation enthalpy barrier decreases linearly with the applied shear stress. On the other hand, between 550 and 800 MPa a plateau exists, while above 800 MPa the activation enthalpy decreases slowly. Therefore, it is not possible to fit the activation enthalpy with a single Kocks law. These results suggest a transition in the basal slip mechanisms at different stress regimes. This transition will be shown in the next section to be directly related to the shape and atomic structure of the critical kink nucleus in the different stress regimes.

The enthalpy versus applied stress as predicted from the MD simulations at different stresses are also shown in Figure \ref{Fig:Delta_H_vs_tau}. A clear discrepancy between the NEB and MD predictions is observed, indicating that the predicted basal slip mechanism is also not the same in the NEB and MD simulations.
As mentioned above, in the MD simulations the dislocation is switching its core configuration to a prismatic-I core while transferring from one pyramidal-I plane to another. To assess the validity of the activation enthalpies extracted from MD simulations, another set of NEB calculations were performed with two prismatic-I core on the same basal plane separated by one Peierls barrier as shown by the green line in the inset of Figure \ref{Fig:Delta_H_vs_tau}. In this case, the dislocation is oblige to pass through the pyramidal-I plane. The variation of the activation enthalpy with the applied stress is shown in green symbol in Figure \ref{Fig:Delta_H_vs_tau}. As expected the dislocation glide is composite combining the pyramidal-I and prismatic-I slips. At first, the dislocation proceeds through the nucleation of a kink pair on the pyramidal-I plane. Then, the dislocation will glide to the ground state pyramidal-I plane, after which it will cross slip back to a prismatic-I core located on the same prismatic-I plane as the center of the dislocation in the final Peierls valley marked by red cross. Finally, an easy prismatic-I slip leads the prismatic-I core to the final position on the basal plane. Interestingly, the enthalpy barriers for such mechanism are very close to those obtained from MD simulations. The difference can be due to thermal effects, since NEB calculations are performed at 0K while MD simulations above 400K.

\subsection{Critical nucleus and atomic structure of the kinks}

\begin{figure}
	\centering
	\includegraphics[width= \linewidth]{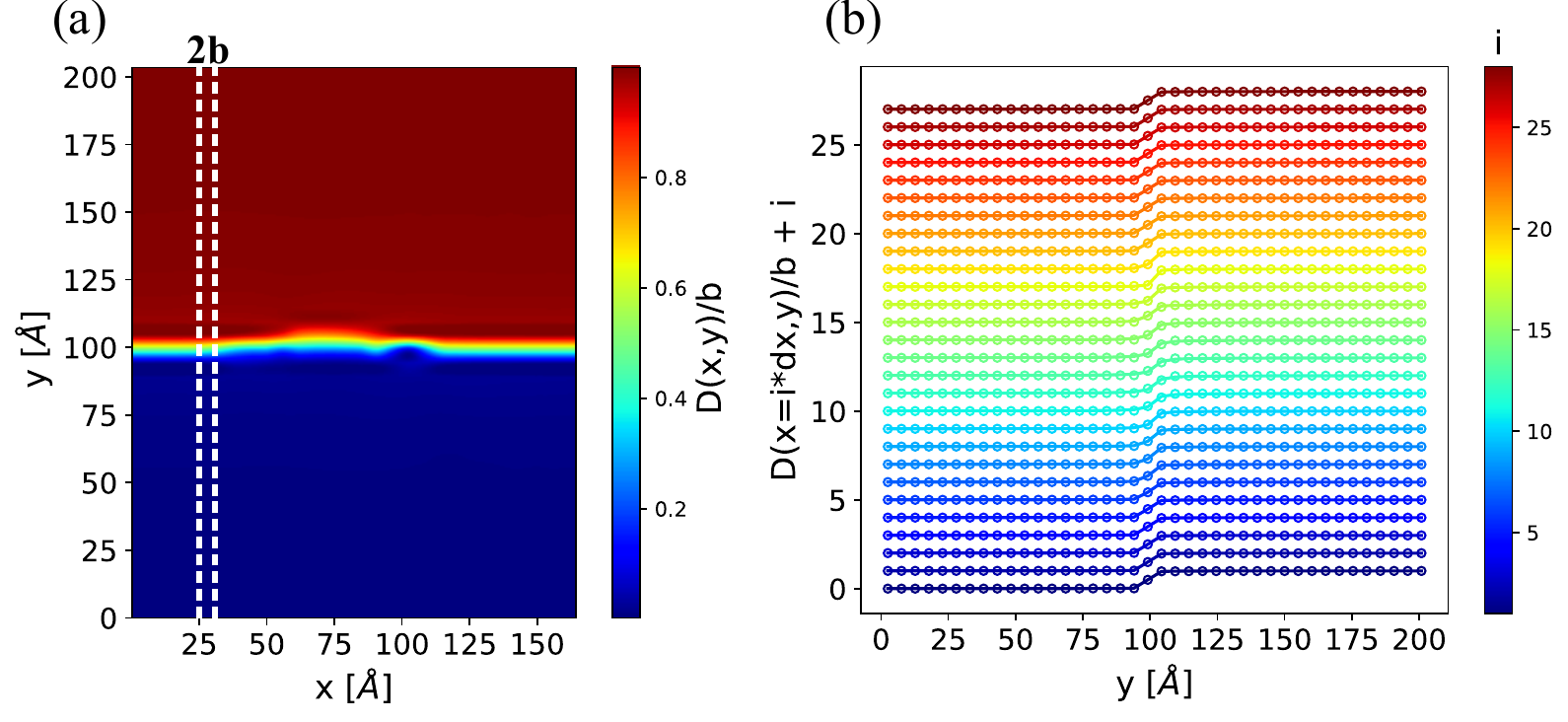}
	\caption{(a) Disregistry created by the kinked dislocation on the glide plane at the highest enthalpy and with zero applied shear stress. (b) Fit of the Peierls Nabarro model shown in colored solid lines onto the disregistry data in intervals $i$ of width $dx =2b$ along the dislocation line shown in colored symbols to determine the position of the dislocation center corresponding to $D(x=i*dx,y) = 0.5$. The disregistry data and their fit were shifted by their index position, $i$, along the dislocation line for clarity of the plot.}
	\label{Fig:Disregistry}
\end{figure}

The variation of the enthalpy barrier with the applied shear stress shown in Figure \ref{Fig:Delta_H_vs_tau}, suggests a transition in the basal slip mechanisms in the different stress regimes.    
To understand this transition, the shape of the critical nucleus corresponding to the highest enthalpy at each stress is characterized here. The shape of the dislocation line at the highest enthalpy was determined from the disregistry $D(x,y)$ created by the kinked dislocation on the basal plane. The disregistry is calculated as:
\begin{equation}
    D(x,y) = U^{+}(x,y) - U^{-}(x,y)
\end{equation}
where $U^{+}(x,y)$, and $U^{-}(x,y)$ are the displacement for atoms on the basal planes just above and below the glide plane, respectively. Figure \ref{Fig:Disregistry}(a) shows the disregistry plot of the dislocation line shape at the highest enthalpy and with zero applied stress. The Pierels Nabarro model for a compact dislocation is then fitted to the disregistry plot in intervals of width $dx = 2b$ to determine the $y(x)$ variation in the dislocation line shape. The fit of the Pierels Nabarro model to the disregistry plot of the dislocation is shown in Figure \ref{Fig:Disregistry}(b). The Peierls Nabarro model is observed to perfectly fit the dislocation disregistry data enabling an accurate description of the shape of the dislocation. 

\begin{figure}
	\centering
	\includegraphics[width= \linewidth]{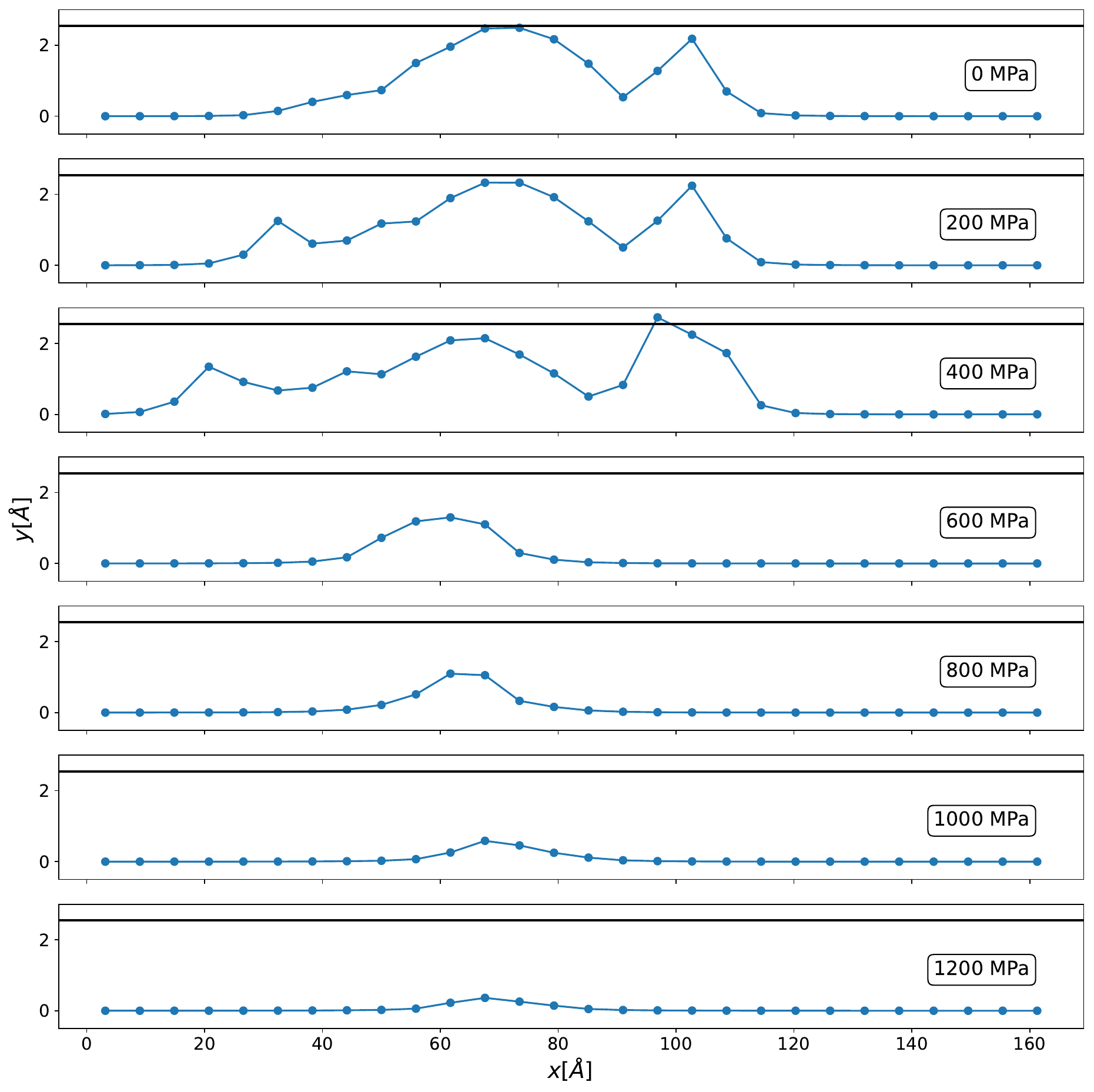}
	\caption{The dislocation profiles for the critical  nucleus at different applied shear stresses.}
	\label{Fig:Dislocation_profiles}
\end{figure}

Figure \ref{Fig:Dislocation_profiles} shows the dislocation profile for the critical nucleus at different applied stresses. In the first stress regime below 550 MPa, the critical nucleus extends all the way from the first Peierls valley to the second and become wider with the increase of the applied stress. On the other hand, above 550 MPa the kink height decreases with the applied stress and it is not elongated all the way from the first valley to the second as in the prior case. Additionally, the dislocation line at high applied stress becomes smoother and converges to an almost straight line at very high stresses (cf. 1200 MPa). This clear transition in the shape of the critical  nucleus confirms the transition in the slip mechanisms with the increase in the applied stress, which was indicated by the different Kocks law fits shown in Figure \ref{Fig:Delta_H_vs_tau}. This dependency of the critical bulge or  nucleus on the applied stress was commonly described by line tension models \cite{caillard2003thermally}. At high stresses the variation in the line energy during the bulge formation is much more important than the variation of the elastic interaction between the kinks forming the bulge \cite{caillard2003thermally}. The opposite is true at low stresses. Starting from the higher stresses, the critical bulge approximation becomes invalid when
the leading part of the critical bulge approaches the bottom of the neighbouring valley and tends to straighten as shown in the low stress regime. From the line tension approximation, this transition happens around $\tau= \Delta E_p/(b \lambda_p)$, where $\Delta E_p$ is the height of the Peierls barrier. Using the value of $\Delta E_p \approx 17$ meV/$\AA$, as shown in Figure \ref{Fig:basal_pyramidal_slip_barrier}, then $\tau_{xz} = 365$ MPa. This value is slightly lower than the 500-600 MPa range for the transition to happen from the NEB calculations observed.

\begin{figure}
	\centering
	\includegraphics[width= 0.6 \linewidth]{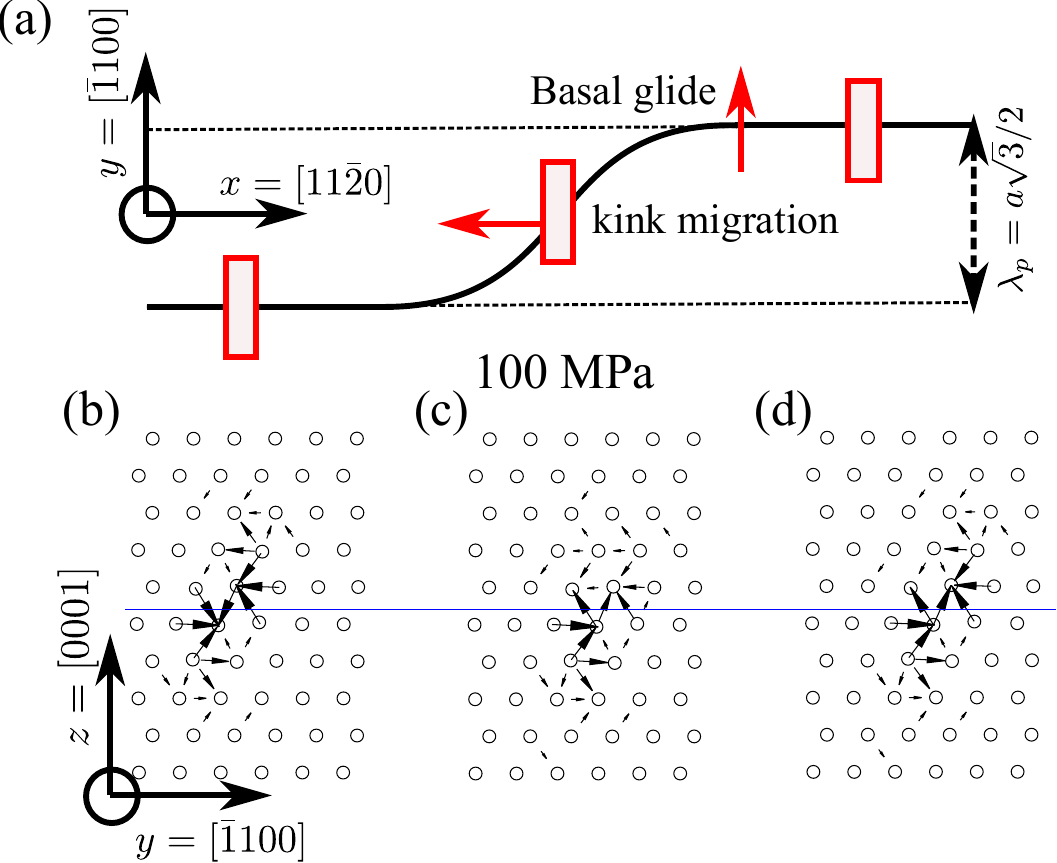}
	\caption{(a) A schematic of the shape of the dislocation line in the low stress regime at 100 MPa. The differential displacement plots at the (b) initial Peierls valley; (c) the kinked part of the dislocation; and (d) the final Peierls valley are also shown, which locations are shown in (a) by the red boxes from left to right, respectively. The blue line in these plots indicates the basal glide plane.}
	\label{Fig:low_stress_regime}
\end{figure}

The atomic structure of the critical  nucleus was also investigated at different applied shear stresses. Figure \ref{Fig:low_stress_regime} shows the atomic structure of the critical nucleus in the low stress regime at 100 MPa using differential displacement plots at different positions along the dislocation line. In this low stress regime the critical  nucleus extends from the first to the next Peierls valley as shown in Figure \ref{Fig:Dislocation_profiles}. In the initial and final Peierls vallies, the core adopts its ground state configuration, which is dissociated on the pyramidal-I plane separated by one Peierls distance $\lambda_p$ on the basal plane. However, in the kinked part, the dislocation exhibit an atomic configuration with a mixture of basal and prismatic dissociation with its center lying in the same basal plane as the atomic configurations at both Peierls valleys. Therefore, the kinked part of the dislocation is lying on the basal plane, and thus, dislocation glide is fully resolved in the basal plane. This is in agreement with experimental observations where basal slip was proposed to be the elementary slip system \cite{caillard2018glide}.

\begin{figure}
	\centering
	\includegraphics[width= \linewidth]{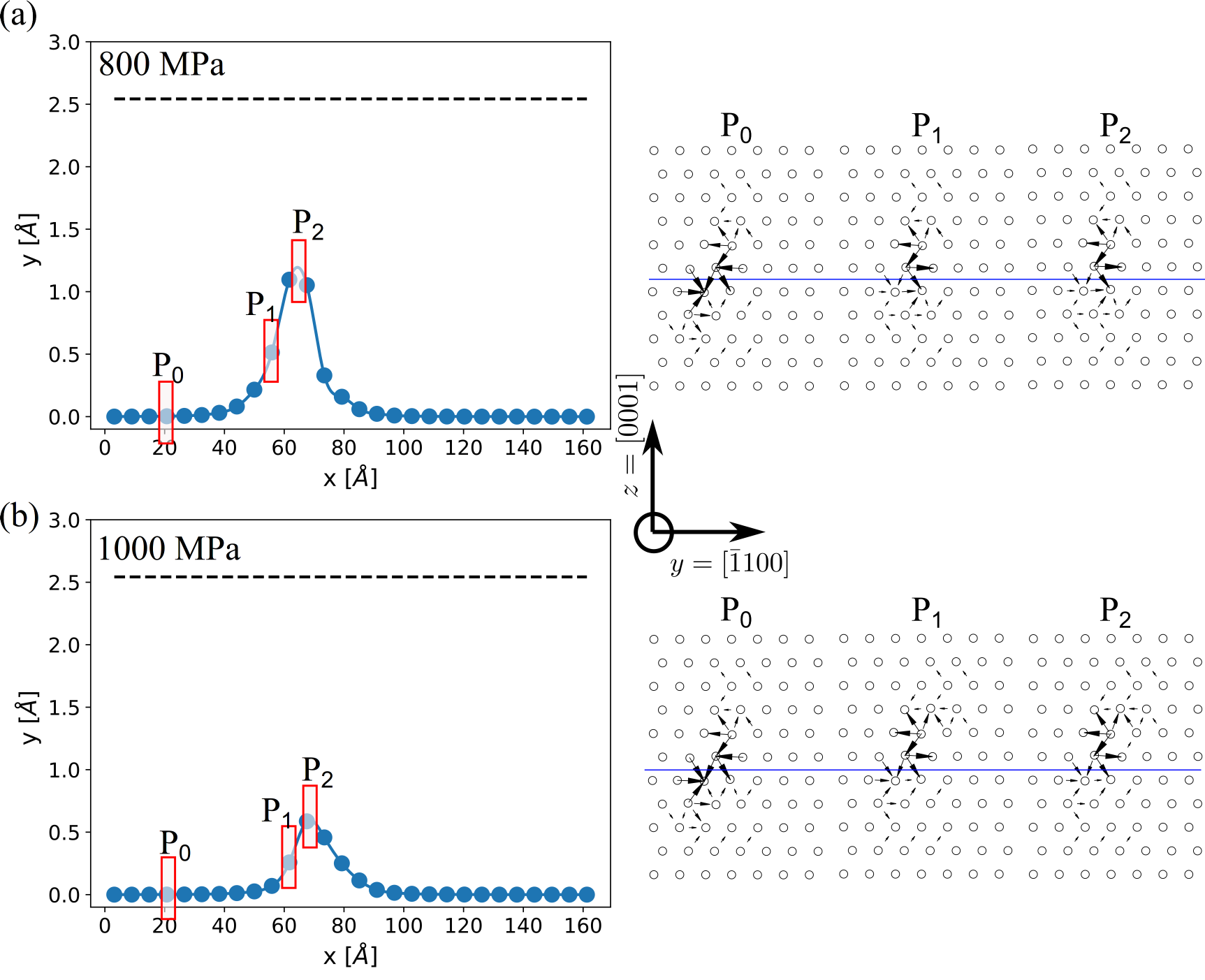}
	\caption{Dislocation profile for the critical nucleus at (a) 800 MPa and (b) 1000 MPa. Differential displacement plots corresponding to different sections along the dislocation line marked by red boxes and named as P$_0$, P$_1$, and P$_2$ are also shown for each stress value. The blue line in these plots indicates the basal glide plane. }
	\label{Fig:high_stress_regime}
\end{figure}

The atomic structure of the dislocation core along the dislocation line in the high stress regime is shown in Figure \ref{Fig:high_stress_regime} at different applied shear stresses, namely, 800 and 1000 MPa. These two stresses were chosen to highlight the transition from the second to the third stress regimes shown in Figure \ref{Fig:Delta_H_vs_tau}. In the initial Peierls valley at both stresses the dislocation core is the ground state pyramidal-I core configuration. At 800 MPa, the dislocation core in the kinked part and the tip of the critical nucleus shows a mixed prismatic and pyramidal spreading, as shown in the differential displacement plots of Figure \ref{Fig:high_stress_regime}(a). This core is different than the unstable core configuration dissociated on the pyramidal-I plane at 1000 MPa (Figure \ref{Fig:high_stress_regime}(b)). This difference in the atomic structure of the critical nucleus explains the change in the activation enthalpy between the second and the third stress regimes.    

\section{Discussion}
\label{Sec:discussion}

In the above, the basal slip of $\langle a \rangle$ screw dislocations in pure Ti was studied by MD simulations and NEB calculations. The MD simulations show a stress/temperature equivalence for basal slip in pure Ti. Effective basal slip was also shown by the MD simulations to be produced by the cross-slip  of the dislocation between two pyramidal-I planes. At low stress and/or high temperature, the trajectory of the dislocation is on average on the basal plane, whereas at high stress and/or low temperature deviation from this average basal slip can be observed. These observations in pure Ti are similar to those reported for pure Zr \cite{maras2022secondary}. However, in Zr, the dislocation fluctuates on the prismatic-I plane before transitioning through one of the pyramidal-I planes from one Peierls valley to another on the basal plane \cite{maras2022secondary}, which is not the case in pure Ti. This Brownian motion of the dislocation on the prismatic-I plane in pure Zr is on one hand related to the absence of friction stress on this plane above room temperature, and on the other hand related to the dissociation of the ground state core configuration of $\langle a \rangle$ screw dislocation on this plane. This latter case is different in pure Ti where the ground state core configuration is dissociated on the pyramidal-I plane, which explain the absence of prismatic fluctuations in the current MD results. It is also worth pointing out that the results in both Ti and Zr are in contradiction with experimental observations which report that basal slip is an elementary slip system \cite{caillard2018glide}. In Zr, this difference was attributed to the higher applied stresses than in experiments \cite{maras2022secondary}. Whereas, in Ti this behavior is related to the cross-slip of the $\langle a \rangle$ screw dislocations from the pyramidal-I core to the prismatic-I core due to thermal fluctuation as shown in the NEB calculations for the prismatic-I core in Figure \ref{Fig:Delta_H_vs_tau}.

In addition to the MD results, the current NEB calculations have also shown that in the low stress regime, basal slip operates through kink-pair nucleation and propagation mechanism, where the kinks nucleate in the basal plane thereby leading to a glide of the screw dislocation fully resolved on this basal plane. Therefore, basal slip is an elementary slip system in pure Ti at low stress. This is in excellent agreement with experiments showing well defined slip traces on basal planes \cite{caillard2018glide}. Akhtar \cite{akhtar1975basal} estimated the activation energy and the temperature dependence of the activation volume for basal slip from tensile stress experiments above 500K. An activation energy of 2.5 eV was found from the fit of a visco-plastic thermally activated law on their tensile test data. This activation energy is slightly higher than the one obtained from the current NEB calculations for zero applied shear stress (2.0 eV). Akhtar proposed that basal slip operates through a cross-slip mechanism due to the high value of the activation energy. On the other hand, as noted by Maras and Clouet in pure Zr, such mechanism necessitates a metastable configuration which can easily glide in basal planes \cite{maras2022secondary}. The best configuration would be a core dissociated in a basal plane, but ab-initio calculations have shown that such core is unstable, which is also reproduced by the Ehemann potential. Therefore, as in the Zr case \cite{akhtar1973basal, maras2022secondary} the high value of the activation energy and the activation volume found by Akhtar are more likely the signature of another mechanism limiting basal glide at high temperatures.

In addition, the atomic structure of the kinks in pure Ti in the low stress regime (Figure \ref{Fig:low_stress_regime}) is found to be similar to that of in pure Zr \cite{maras2022secondary}. The only exception is that the dislocation is dissociated on the ground state pyramidal-I plane and not on the prismatic-I plane in its initial and final Peierls valley. However, thermal activation can lead to a cross-slip of the core of the dislocation in both Peierls valley to a highly mobile prismatic-I plane, thereby leading to wavy slip traces as observed experimentally \cite{akhtar1975basal, caillard2018glide}.

\section{Conclusion}
 Molecular dynamics (MD) simulations and nudged elastic band (NEB) calculations were performed to study the mechanism of basal slip of $\langle a \rangle$ screw dislocations in pure Ti. Our MD results showed that basal and pyramidal slip are intertwinned. The screw dislocation glides by kink pair nucleation and propagation mechanism on one of the two equiprobable pyramidal planes adjacent to the maximum resolved shear stress basal plane, before switching to the other pyramidal plane. At low stress and/or high temperature the gliding distance on each pyramidal-I plane is short leading to an average basal slip. However, at high stress and/or low temperature the average slip plane is shifted from the basal plane due to the longer gliding distance on each pyramidal-I plane. In contrast to MD simulations, NEB calculations revealed that at low stresses, kinks are not nucleated anymore in the pyramidal-I plane but rather directly in the basal plane fully compatible with the motion of a screw dislocation confined to the basal plane in agreement with experimental observations in Ti and Zr \cite{caillard2018glide, maras2022secondary}. The discrepancy between these two mechanisms is shown to be related to the thermal activation of the cross-slip from the pyramidal-I to the prismatic-I core configuration of the $\langle a \rangle$ screw dislocation.

\section*{Acknowledgements}
This research was sponsored by the Multiscale Structural Mechanics and Prognosis program at the Air Force Office of Scientific Research (project number: FA9550-21-1-0028). Some simulations were conducted at the Advanced Research Computing at Hopkins (ARCH) core facility (rockfish.jhu.edu), which is supported by an NSF grant number OAC-1920103. Some simulations were also conducted using the Extreme Science and Engineering Discovery Environment (XSEDE) Expanse supercomputer at the San Diego Supercomputer Center (SDSC) through allocations TG-MAT210003. XSEDE is supported by National Science Foundation grant number ACI-1548562.
 
\bibliography{biblio} 

\end{document}